# Decentralized Privacy-Preserving Proximity Tracing

Version: 25 May 2020.
Contact the first author for the latest version.

**EPFL**: Prof. Carmela Troncoso, Prof. Mathias Payer, Prof. Jean-Pierre Hubaux, Prof. Marcel Salathé, Prof. James Larus, Prof. Edouard Bugnion, Dr. Wouter Lueks, Theresa Stadler, Dr. Apostolos Pyrgelis, Dr. Daniele Antonioli, Ludovic Barman, Sylvain Chatel

**ETHZ**: Prof. Kenneth Paterson, Prof. Srdjan Čapkun, Prof. David Basin, Dr. Jan Beutel, Dr. Dennis Jackson, Dr. Marc Roeschlin, Patrick Leu

**KU Leuven**: Prof. Bart Preneel, Prof. Nigel Smart, Dr. Aysajan Abidin

**TU Delft**: Prof. Seda Gürses

**University College London**: Dr. Michael Veale

**CISPA**: Prof. Cas Cremers, Prof. Michael Backes, Dr. Nils Ole Tippenhauer

**University of Oxford**: Dr. Reuben Binns

**University of Torino / ISI Foundation**: Prof. Ciro Cattuto

**Aix Marseille Univ, Université de Toulon, CNRS, CPT**: Dr. Alain Barrat

**IMDEA Software Institute**: Prof. Dario Fiore

**INESC TEC**: Prof. Manuel Barbosa (FCUP), Prof. Rui Oliveira (UMinho), Prof. José Pereira (UMinho)

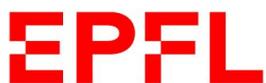 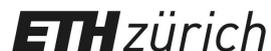 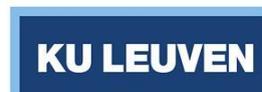 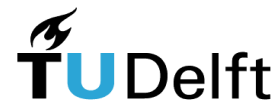

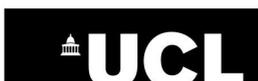 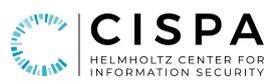 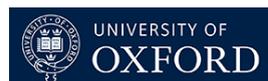 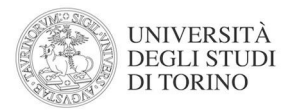

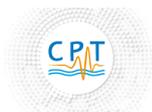 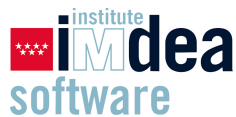 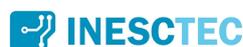

## Executive Summary

This document describes and analyzes a system for secure and privacy-preserving proximity tracing at large scale. This system provides a technological foundation to help slow the spread of SARS-CoV-2 by simplifying and accelerating the process of notifying people who might have been exposed to the virus so that they can take appropriate measures to break its transmission chain. The system aims to minimise privacy and security risks for individuals and communities and guarantee the highest level of data protection.

The goal of our proximity tracing system is to determine who has been in close physical proximity to a COVID-19 positive person and thus exposed to the virus, without revealing the contact's identity or where the contact occurred. To achieve this goal, users run a smartphone app that continually broadcasts an ephemeral, pseudo-random ID representing the user's phone and also records the pseudo-random IDs observed from smartphones in close proximity. When a patient is diagnosed with COVID-19, she can upload pseudo-random IDs previously broadcast from her phone to a central server. Prior to the upload, all data remains exclusively on the user's phone. Other users' apps can use data from the server to locally estimate whether the device's owner was exposed to the virus through close-range physical proximity to a COVID-19 positive person who has uploaded their data. In case the app detects a high risk, it will inform the user.

The system provides the following security and privacy protections:

- **Ensures data minimization**. The central server only observes anonymous identifiers of COVID-19 positive users without any proximity information. Health authorities learn no information except that provided when a user reaches out to them after being notified.

- **Prevents abuse of data**. As the central server receives the minimum amount of information tailored to its requirements, it can neither misuse the collected data for other purposes, nor can it be coerced or subpoenaed to make other data available.

- **Prevents tracking of users.** No entity can track users that have *not* reported a positive diagnosis. Depending on the implementation chosen, others can only track COVID-19 positive users in a small geographical region limited by their capability to deploy infrastructure that can receive broadcasted Bluetooth beacons.

- **Graceful dismantling.** The system will dismantle itself after the end of the epidemic. COVID-19 positive users will stop uploading their data to the central server, and people will stop using the app. Data on the server and in the apps is removed after 14 days.

We are publishing this document to inform the discussion revolving around the design and implementation of proximity tracing systems. This document is accompanied by other documents containing an overview of the data protection compliance of the design, an extensive privacy and security risk evaluation of digital proximity tracing systems, a proposal for interoperability of multiple systems deployed in different geographical regions, and alternatives for developing secure upload authorisation mechanisms.

## Changelog

**25 May 2020**
General:
- Describe context and need for proximity tracing systems.
- **Remove secondary purpose of data sharing for research purposes** (its privacy implications will be discussed in a separate document).
  - Includes clarification on why we think this is a secondary goal.
  - Includes removing epidemiologists from the threat model.

Designs:
- Rename parameter that defines length of epoch from `l` to `L`.
- Smartphones now store information about received beacons rather than per-`EphID` aggregates.
- Clarified why the unlinkable design uploads the seed rather than the resulting `EphID`.
- Removed **authorisation** description and added pointer to "DP3T - Upload Authorisation Analysis and Guidelines".
- Added **one decentralized scheme for better protection against eavesdropping attacks**.
- Added a separate section on **interoperability** including a summary of the document "Interoperability of decentralized proximity tracing systems across regions".
- Added a separate section to describe notification procedure.

Security:
- Improvements to security and privacy leakage analysis in accordance with in-depth analysis on "Privacy and Security Attacks on Digital Proximity Tracing Systems" document.

Comparison:
- Rewritten comparison to centralized designs to compare with concrete systems.

Many of these changes were inspired by the large amount of feedback we received on GitHub. The amount of issues is too long to enumerate, but we want to acknowledge the effort that the community has put into improving this work.

**12 April 2020**
General:
- Add enhancement for decentralised proximity tracing solutions (Section 4) to prevent short-term and remote eavesdropping

Designs:
- Clarified different time units in which system operates for both designs
- Clarified what are configurable system parameters for both designs

**10 April 2020**
General:
- Fixed various typos

- Introduce the two designs early in Section 1

Designs:
- Removed two more mentions of using health care provider to do notification (response issues #52 & #106; Section 2, Interoperability, and Section 4.2)
- Fixed bug in unlinkable design to add correct value to the Cuckoo filter (response issue #77; Section 3, Decentralized proximity tracing)
- Clarify that the notification mechanism (content of the messages, how many repeats, etc.) needs to be further worked out. (Section 2, Notification of risk)
- Added local storage requirement (Section 2 and Section 3, Local storage of observed events)
- After consulting with epidemiologists, we decided that a coarse time indication in days (instead of parts of days) provides sufficient granularity. (Updated throughout)
- Removed storing of auxiliary data with observed EphIDs from all designs as it is currently not required (Section 2, 3 and 5)
- Give concrete example of false positives for the Cuckoo filter (Section 3)

Security and Privacy analysis:
- Corrected suggestion that traffic analysis works against at-risk individuals. (Section 4.2)
- Clarified that in the unlinkable design it is also impossible to claim another user's EphID as one's own. (Section 4.3)

**7 April 2020**
General:
- Numbered sections for easier referencing.

Goals and requirements:
- Clarify app sends notification (Section 2)
- Add detail about most relevant information for epidemiological analysis (Section 1.1)
- Add non-goals of the system (response issue #33, Section 1.1)

Previous design (renamed: **Low-cost decentralized proximity tracing,** in Section 2)
- Clarification on data sent to epidemiologists (Section 2, Epidemiologists)
- Slight tweak to design: send the day $t$ explicitly when reporting an infected key $SK_t$, be clear that $t$ is a global rather than local counter (Section 2, Setup)
- Added second operation point in scalability (Section 2, Scalability)
- Interoperability: added possibility of mobile network carrier, and changed hard coded to config file (response issue #26, Section 2, Interoperability)

**Added alternative design (Unlinkable decentralized proximity tracing**, in Section 3)
- Added a new design that: prevents broadcast of seeds, provides unlinkability between EphIDs, enables users to redact EphIDs that they do not want published

Security and Privacy analysis
- Clarified that no notification is automatically sent to the health authority (response issue #52)
- Added introduction to threat model (response issue #47, Section 4.1)
- Removed redacting mitigation from low-cost design (not possible due to hash chain)
- Added replay attack to create fake contacts on low-cost design (Section 4.3)
- Added security and privacy analysis of unlinkable design (Sections 4.2 and 4.3)
- Added comparison with both unlinkable design and low-cost design in the table (Section 5.4)

**3 April 2020**
Initial public release

# Contents

## 1. Need, purpose and requirements

In this section, we describe the problems that motivate the need for digital proximity tracing systems, the purpose of our system, and its requirements. We discuss additional, potentially desirable goals that have been proposed for digital proximity tracing that our design does not attempt to achieve.

In the next sections, we present three protocols to implement decentralized proximity tracing. One protocol is an extremely lightweight system that permits limited tracing of COVID-19 positive users under very specific conditions. The other protocols provide additional privacy properties at a moderate increase in cost.

## 1.1 Context and need

Beyond its medical and economic consequences, the COVID-19 pandemic poses a severe challenge to healthcare authorities and governments: how to contain the spread of the SARS-CoV-2 virus while simultaneously returning to normality. There has been a vigorous debate about the best strategy to achieve these two goals. However, many experts advocate for a strategy based on testing, contact tracing, isolation, and quarantine[1] (TTIQ - see also the contact tracing[2] and proximity tracing[3] policy briefs by the Swiss National COVID-19 Science Task Force).

A cornerstone of the TTIQ strategy is effective contact tracing. Contact tracing **identifies individuals who have been exposed** to a COVID-19 positive person and consequently are at risk of having contracted COVID-19. Identifying the contacts of a confirmed positive case, so they can go into quarantine as quickly as possible**,** is of crucial importance to successfully containing the spread of the virus. In particular, presymptomatic transmission - i.e., transmission during the 2-3 days before the onset of symptoms - is estimated to account for about half of the overall transmission.[4] Thus, asking all exposed contacts to go into quarantine very rapidly is key in breaking the transmission chains of the virus.

Manual contact tracing relies on interviews conducted by trained personnel. This process alone is limited in responding to the demands of COVID-19 for two reasons:

1) In-person or over-the-phone contact tracing interviews are time consuming, and in order to handle a large number of infected people, they require many trained

---

[1] Salathé M et al. COVID-19 epidemic in Switzerland: on the importance of testing, contact tracing and isolation. *Swiss Med Wkly.* March 19, 2020

[2] Swiss National COVID-19 Science Task Force. "SARS-CoV-2 contact tracing strategy: epidemiologic and strategic considerations" (26 April 2020) Accessed on 23 May 2020:
https://ncs-tf.ch/en/policy-briefs/contact-tracing-strategy-26-april-20-en/download

[3] National COVID-19 Science Task Force. "Digital Proximity Tracing" (15 May 2020) https://ncs-tf.ch/en/policy-briefs/digital-proximity-tracing-15-may-20-en/download

[4] See e.g. He X et al. Temporal dynamics in viral shedding and transmissibility of COVID-19. Nature Medicine 2020; Ganyani T et al. Estimating the generation interval for coronavirus disease (COVID-19) based on symptom onset data, March 2020. Eurosurveillance 2020.

contact tracers and are therefore difficult to scale rapidly.

2) Even with a long, in-depth interview, the list of contacts from the interview is often incomplete. In the case of a respiratory disease, such as COVID-19, any person who has been in close physical proximity to a COVID-19 positive person should be listed as a contact. This includes strangers that a diagnosed person will not be able to recall or identify in an interview, such as nearby passengers on public transportation. Even remembering all acquaintances one has encountered over the past two weeks can be a challenge.

These issues have prompted numerous initiatives towards the use of digital proximity tracing systems to support human contact tracers.

## 1.2 Purpose

The primary purpose of digital proximity tracing systems is to provide a **mechanism that alerts users who have been in close physical proximity to a confirmed COVID-19 positive case for a prolonged duration** that they may have been exposed to the virus. Exposure to the virus does not imply that the person has contracted COVID-19, but serves as a trigger for a precautionary intervention recommended by a public health authority, such as testing or quarantine. This process does not require revealing the identity of the diagnosed person or when and where the contact occurred.

Most adults carry smartphones throughout the day. This opens the possibility of a mobile application that collects data about close physical proximity between individuals and thus allows the tracing of contacts that might have been infected through droplet transmission, widely assumed to be the dominant transmission route of SARS-CoV-2[5]. To this end, the application records exposure events between personal smartphones. An **exposure event** is recorded when two phones are in close physical proximity for a period of time, for some pre-defined value for distance and duration. **Proximity tracing** is the process that the app uses to calculate whom to notify of a high-risk exposure..

Digital proximity tracing **is a complement, not a substitute, for manual contact tracing**. It can augment the efforts of health officials, gaining precious time as alerts can be sent automatically and can inform otherwise unidentifiable contacts of a COVID-19 positive person.

### *Non-goals*

Our system does not aim to achieve the following goals:

- **Track positive cases**: The system does neither attempt to provide a mechanism to track users who report a positive COVID-19 diagnosis through the app, nor to ensure that confirmed positive cases comply with medical orders. We assume that users who have received a positive test result act responsibly and take necessary precautions. Therefore, we do not attempt to detect contacts of confirmed positive cases *after* their diagnosis nor do we attempt to detect or prevent misbehavior. In our view, the gain in utility (potentially detecting irresponsible behavior of few individuals) does not justify the loss of privacy for the majority of users who adhere to guidelines to protect their fellow citizens. Moreover, our system does not provide location-tracking functionality and cannot determine when a user is "in public."

---

[5] US CDC How COVID-19 spreads:
https://www.cdc.gov/coronavirus/2019-ncov/prevent-getting-sick/how-covid-spreads.html

- **Detect hotspots or trajectories of positive cases**: The system does not attempt to identify locations frequented by confirmed positive cases, which might increase others' risk of exposure. This is a deliberate design decision. We limit the purpose of our system to its primary goal. This choice enables us to collect and process very little data. In particular it avoids collecting location data, which is highly sensitive and very difficult to publish in a privacy-preserving way.

- **Sharing data for research purposes**[6]: The system is not designed to support epidemiological research. As a side effect of fulfilling their primary purpose, proximity tracing systems produce data about close-range proximity between personal smartphones that might be of great value to epidemiologists and related research groups. This has triggered a public debate about whether proximity tracing systems should be designed specifically to collect additional data that might help epidemiologists improve their understanding of and predictions about the spread of SARS-CoV-2.

  We strongly believe, however, that it is not the time to conflate novel, untested technologies with the understandable desire to collect new epidemiological insights. Furthermore, the data collected by proximity tracing applications does **not** allow inferences about causal transmission chains (who infected whom), fomite transmission (transmission through surfaces of objects), or aerosol transmission (transmission via aerosols that remain suspended in the air for some time). We thus designed a system that is optimised to fulfill its primary purpose and to support and complement manual contact tracing through measurement of proximity over a prolonged period of time. How much of this data should be shared to support epidemiological research purposes is a separate question. We plan to publish a separate analysis of the privacy implications of data sharing.

---

[6] It is in theory possible for proximity tracing systems to additionally share data intended for research purposes. However, this would invalidate the existing security and privacy analyses, and would require additional in-depth investigations into the impact of the shared data and the interaction with the other functions of the system.

## 1.3 System requirements

### 1) Enable proximity tracing

To fulfill its primary purpose, the application must provide the following properties:

- **Completeness:** The exposure history captures all exposure events.

- **Precision:** Reported exposure events must reflect actual physical proximity.

- **Authenticity:** Exposure events are authentic, i.e., users cannot fake exposure events.

- **Confidentiality:** A malicious actor cannot access the contact history of a user.

- **Notification:** Individuals can be informed about prolonged exposure to the virus.

### 2) Respect and preserve digital right to privacy of individuals

It is of paramount importance that any digital solution to proximity tracing **respects the privacy of individual users and communities** and **complies with relevant data protection guidelines** such as the European General Data Protection Regulation (EDPB Statement on GDPR and COVID-19) or the related Swiss law. The GDPR does not prevent the use of personal data for public health, particularly in times of crisis, but it still imposes a binding obligation to ensure that 'only personal data which are necessary for each specific purpose of the processing are processed' (art 25). It is therefore a legal requirement to consider, particularly in the creation of systems with major implications for rights and freedoms, whether such a system could be technically designed to use and retain less data while achieving the same effect. To this end, an application must minimize the amount of data collected and processed to avoid risks for individuals and communities, and it should reveal only the minimum information truly needed to each authorized entity.

Furthermore, a common concern with systems such as these is that the data and infrastructure might be used beyond its originally intended purpose. Data protection law supports the overarching principle of '**purpose limitation**' — precluding the widening of purposes after the crisis through technical limitations. Such assurances will likely be important to achieve the necessary level of adoption in each country and across Europe, by providing citizens with the confidence and trust that their personal data is protected and used appropriately and carefully. Only applications that do not violate a user's privacy **by design** are likely to be widely accepted.

The system should provide the following guarantees:

- **Data use:** Data collection and use should be limited to the purpose of the data collection: proximity tracing. This implies that the design should avoid collecting and using any data, for example geolocation data, that is not directly related to the task of detecting a close contact between two people.

- **Controlled inference:** Inferences about individuals and communities, such as information about social interactions or medical diagnosis, should be controlled to avoid unintended information leakage. Each authorised entity should only be able to learn the information strictly necessary to fulfill its own requirements.

- **Protect identities:** The system should collect, store, and use anonymous or pseudonymous data that is not directly linkable to an individual's identity where possible.

- **Erasure:** The system should respect best practices in terms of data retention periods and delete any data that is not relevant.

## 3) Fulfill the scalability requirements posed by a global pandemic

SARS-CoV-2 is rapidly spreading across the globe following people across national borders and continents. As a core principle of free democratic societies, after the current confinement measures end, free movement should resume. Proximity tracing must support free movement across borders and scale to the world's population.

The system should provide the following guarantees:

- **Scalability:** The system scales to billions of users.

- **Interoperability:** The system works across borders and health authorities.

## 4) Feasibility under current technical constraints

There is an urgency to not only design but ***implement*** a digital system that simplifies and accelerates proximity tracing in the near future. This mandates a system design that is ***mindful of the technical constraints*** posed by currently available technologies.

- **No reliance on new breakthroughs:** The system should, as far as possible, only use techniques, infrastructure, and methods readily available at the time of development and avoid relying on new breakthroughs in areas such as cryptography, GPS localisation, Bluetooth or Ultra Wide Band distance measurements; or new deployments such as novel anonymous communications systems that have not been widely tested for privacy.

- **Widely available hardware**: The goal of high adoption of proximity tracing can only be achieved if both server- and client-side applications can run on widely available smartphones and server hardware.

## 2. Decentralized proximity tracing

We propose a privacy-friendly, decentralized proximity tracing system that reveals minimal information to the backend server. We propose three different protocols to support exposure detection and tracing. These protocols provide developers with choice regarding the trade-off between privacy and computation cost but share a common framework.

In all three protocols, smartphones locally generate frequently-changing *ephemeral identifiers* (`EphIDs`) and broadcast them via Bluetooth Low Energy (BLE) beacons. Other smartphones observe these beacons and store them together with a time indication and measurements to estimate exposure (e.g., signal attenuations). See Figure AA.

The proximity tracing process is supported by a *backend server* that distributes anonymous exposure information to the app running on each phone.[7] This backend server is trusted to not add information (i.e., to not add fake exposure events) nor remove information (i.e., to not remove exposure events) and to be available. The backend acts **solely** as a communication platform and does not perform any processing. It is **untrusted with regards to protecting users' privacy**. In other words, the privacy of the users in the system does not depend on the actions of this server. Even if the server is compromised or seized, their privacy remains intact.

If patients are diagnosed with COVID-19, they will be authorized by health authorities to publish a protocol-specific representation of their `EphIDs` for the contagious period to aid in decentralized proximity tracing. We are aware that each country, and in some cases each country's regions, will have existing processes and systems in place to manage mass testing, to communicate between testing facilities and laboratories, and to inform patients. In a separate document,[8] we discuss three proposals for secure mechanisms to validate upload requests from personal devices to the central backend and evaluate their usability trade-offs. Here, we leave the authorisation mechanism abstract. We further note that some implementations of the system might skip the authorisation step altogether and rely on self-reporting. However, we strongly advise implementing one of the proposed authorisation mechanisms to achieve stronger security guarantees.

When authorized, users can instruct their phones to upload a representation of the `EphIDs` to the backend. The backend stores the uploaded *representations*. To protect

---

[7] We assume that the MAC address of the phone changes every time the ephemeral identifier (EphID) changes to prevent prolonged tracking of smartphones.

[8] "Secure Upload Authorisation for Digital Proximity Tracing", The DP-3T Project, https://github.com/DP-3T/documents/blob/master/DP3T%20-%20Upload%20Authorisation%20Analysis%20and%20Guidelines.pdf

COVID-positive users from network observers, all phones equipped with the app generate dummy traffic to provide plausible deniability of real uploads.

Other smartphones periodically query the backend for information and reconstruct the corresponding `EphIDs` of COVID-19 positive users locally. If the smartphone has recorded beacons corresponding to any of the reported `EphIDs`, then the smartphone's user might have been exposed to the virus. The smartphone uses the exposure measurements of the matched beacons to estimate the exposure of the phone's owner, see Section 4.

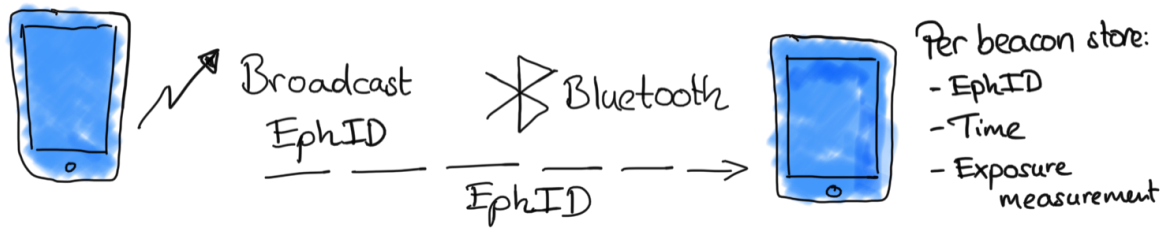

**Figure AA: Processing and storing of observed beacons.**

## 2.1 Low-cost decentralized proximity tracing

In this section, we present a low-cost protocol that has good privacy properties and very small bandwidth requirements.

### Setup

*Initial seed generation.* Let `t` be the current day. Smartphones generate a random initial daily seed `SK`$_t$ for the current day `t`. We assume days correspond to UTC days.

### Creating ephemeral IDs (`EphIDs`)

*`EphID` Generation*. Each day, smartphones rotate their secret day seed `SK`$_t$ by computing

$$SK_t = H( SK_{t-1} ),$$

where `H` is a cryptographic hash function. The smartphone will use the seed `SK`$_t$ during day `t` to generate `EphIDs`.

To avoid location tracking via broadcast identifiers, devices should frequently change the ephemeral identifier `EphID` that they broadcast to other devices. We refer to the duration for which a device broadcasts the same `EphID` as an **epoch**. The length of an epoch, in minutes, is a configurable system parameter that we denote as `L`.

At the beginning of each day `t`, smartphones locally generate a list of `n = (24 * 60)/L` new `EphID`$_i$s to broadcast during day `t`. Given the day seed `SK`$_t$, each device computes

$$EphID_1 \mathbin{||} ... \mathbin{||} EphID_n = PRG( PRF(SK_t, \text{``broadcast key''}) ),$$

where `PRF` is a pseudo-random function (e.g., HMAC-SHA256), `"broadcast key"` is a fixed, public string, and PRG is a pseudorandom generator (e.g. AES in counter mode) producing `n * 16` bytes, which we split into 16-byte chunks to obtain the `n` ephemeral Bluetooth identifiers `EphID` for the day.

Smartphones pick a random order in which to broadcast the `EphID`s during the day. Each `EphID`$_i$ is broadcast for `L` minutes.

## Local storage of observed `EphID`s and seeds $SK_t$

For each received beacon, phones store:

- The received ephemeral Bluetooth identifier `EphID`.

- The exposure measurement (e.g., signal attenuation).

- The day on which this beacon was received (e.g., "April 2").

Note that an `EphID` could be received multiple times and will result in multiple entries in the database (Figure AA). For efficient storage, we propose to group these entries by `EphID`, resulting in 36 bytes per `EphID`. Given a very conservative estimate of 140k different observations over the course of 14 days (i.e., if epochs are 15 minutes, these are 100 different people observed per epoch), this would require around 6.1 MB.

In addition, each device stores the seeds $SK_t$ it generated during the past 14 days. This parameter (i.e., 14 days), which defines the maximum period for which any data (both observed and generated `EphID`s) is stored on the device, is a system parameter and is determined by guidance from health authorities.

## Decentralized proximity tracing

Once the health authority has authorised the proximity tracing for a confirmed COVID-19 positive user (Figure PT, step 1), the user can instruct their phone to send to the backend the seed $SK_t$ and the day $t$ corresponding to the first day in which the user was considered to be contagious (Figure PT, step 2). The start date of the contagious window $t$ can either be determined by the health authority or the user might be trusted to manually enter this day.[9] Epidemiologists estimate that for COVID-19 the contagious window starts 1 to 3 days before the onset of symptoms.

After reporting the seed $SK_t$ and day $t$ for the first day of the contagious window, the smartphone deletes $SK_t$. It then picks a completely new random seed and commences broadcasting `EphID`s derived from this new seed. This ensures that after uploading their past seed, users do not become trackable. Recall that our system does not attempt to track users after reporting a diagnosis because we assume users act responsibly. The new seed will thus only be uploaded if necessary, i.e., if after a second positive diagnosis the user is considered contagious.

---

[9] "Secure Upload Authorisation for Digital Proximity Tracing", The DP-3T Project, https://github.com/DP-3T/documents/blob/master/DP3T%20-%20Upload%20Authorisation%20Analysis%20and%20Guidelines.pdf

Given the seed $SK_t$, everyone can compute all ephemeral identifiers $EphID$ broadcast by the COVID-19 positive user, starting from day $t$ by repeating the process described in [EphID generation](#) above.

The backend collects the pairs $(SK_t, t)$ of COVID-19 positive users. Phones periodically download these pairs (Figure PT, step 3). Each smartphone uses this pair to reconstruct the list of $EphIDs$ of a diagnosed person for each day $t'$ and checks (1) if it has observed any beacon with one of these $EphIDs$ on day $t'$ and (2) that such observations occurred before the corresponding seed $SK_t$ was published.[10] Restricting the matching to a specific day limits replay attacks in which malicious users redistribute captured $EphIDs$ and ensures more efficient lookups.

For each matching recorded beacon (e.g., a beacon with an EphID that was transmitted by a user who reported a positive diagnosis), the beacon's receive time and exposure measurement are taken into account for the exposure risk computation (Figure PT, step 4) and [Section 4](#).

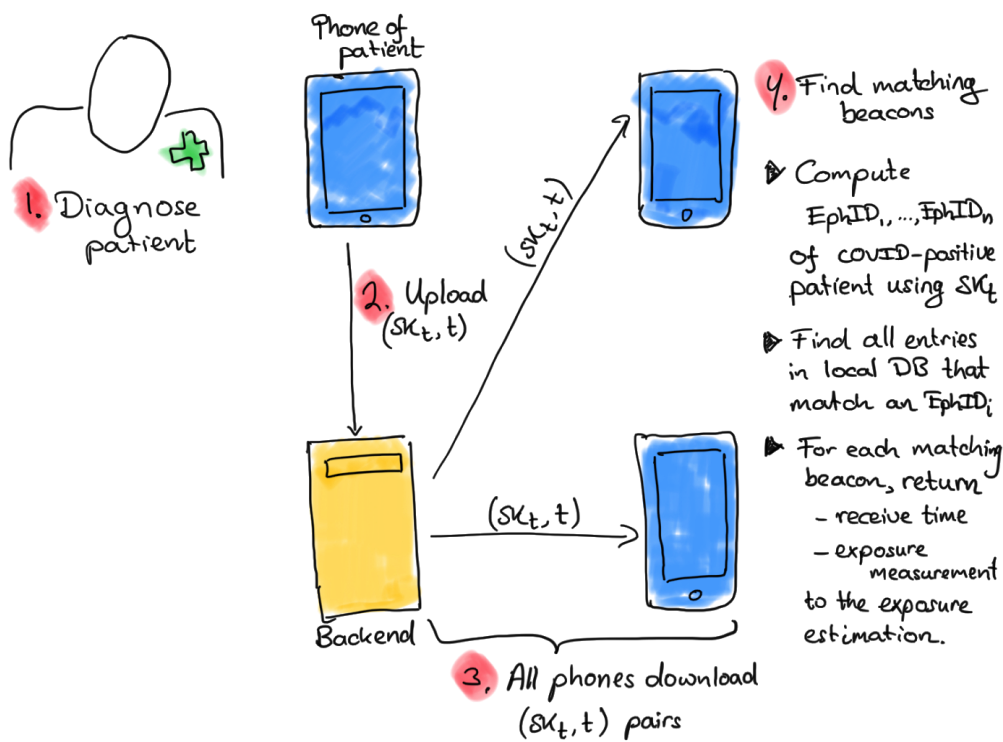

**Figure PT: proximity tracing process.**

## Scalability

For each user who reports a positive diagnosis, the backend needs to store a 32-byte seed and a 2-byte day counter for the duration of the contagious window. Storage cost at the backend is therefore not a problem. Throughout the day, smartphones download the 32-byte seeds and 2-byte day counters of newly diagnosed patients. This data is static and can therefore be effectively served through a content delivery network (CDN).

---

[10] To facilitate this check, the smartphone temporarily stores a more precise receive timestamp of all the beacons it received after the last download from the server. Once all downloads from the server have been processed, the phone coarsens this timestamp for all past observations.

## 2.2. Unlinkable decentralized proximity tracing

In this section, we present a variant of the low-cost design in the previous section that offers better privacy properties than the low-cost design at the cost of increased bandwidth. This design does not disseminate a list containing the seeds of users who have reported a positive diagnosis. Instead, the ephemeral identifiers of COVID-19 positive users are hashed and stored in a Cuckoo filter, which is then distributed to other users.

This design choice offers several advantages. It prevents adversaries from linking the ephemeral identifiers of COVID-19 positive users (see privacy analysis for details), and it enables COVID-19 positive users to redact, after a positive diagnosis, identifiers corresponding to sensitive locations, times, or periods in which users are certain they have not been in contact with other people, e.g. while they were alone or behind a window.

### Setup

No setup is needed.

### Generating ephemeral IDs

As in the low-cost design, smartphones broadcast each ephemeral identifier `EphID` during one epoch of fixed duration `L`. Epochs `i` are encoded relative to a fixed starting point shared by all entities in the system.

Smartphones generate the ephemeral identifier `EphID`$_i$ for each epoch `i` as follows. The smartphone draws a random 32-byte per-epoch seed `seed`$_i$ and sets:

$$\texttt{EphID}_i = \texttt{LEFTMOST128( H( seed}_i \texttt{ ) )},$$

where `H` is a cryptographic hash function, and `LEFTMOST128` takes the leftmost 128 bits of the hash output. Smartphones store the seeds corresponding to all past epochs in the last 14 days. They delete older seeds. As before, the maximum storage period is a system parameter and is determined with guidance from health authorities.

### Local storage of observed EphIDs

For each observed beacon the smartphone stores:

- The hashed string `H(EphID || i)`, where `H` is a cryptographic hash function, and `EphID` the identifier of the beacon, and `i` is the epoch in which the beacon is received. (Note that this differs from the low-cost design, in which the phone stores the raw `EphID`.)

- The exposure measurement (e.g., signal attenuation)

- The day in which this beacon was received (e.g., "April 2").

We include the epoch `i` in the hash to ensure that replaying an `EphID` outside the epoch in which it was originally broadcast can never cause a fake at-risk event (regardless of whether the `EphID` corresponds to a person who later reports a positive diagnosis).

For efficiency of storage, we propose to group these entries by hashed string. A single entry then requires around 52 bytes. Given a very conservative estimate of 140k different observations over the course of 14 days (i.e., 100 people observed per epoch), this would require around 6.9 MB of local storage.

## Decentralized proximity tracing

In case of a positive diagnosis, users can instruct their device to upload a representation of the `EphIDs` produced by the smartphone during the contagious window. Unlike in the low-cost design, the user first has the option to *redact* identifiers by choosing the set `I` of epochs for which they want to reveal their identifiers. For example, the user may want to exclude Monday morning and Friday night. The phone then uploads the set `{(i, seed_i)}` for all epochs `i` in `I`. Requiring `seed_i` rather than the resulting `EphID`, ensures that malicious users cannot claim somebody else's `EphID` as their own (see [security analysis](#)).

Periodically (e.g., every 2 hours), the backend creates a new Cuckoo filter `F` and for each pair `(i, seed_i)` uploaded by a COVID-19 positive user it inserts

$$\text{H ( LEFTMOST128( H( seed}_i \text{ ) ) || i)}$$

into the Cuckoo filter `F`, i.e., the hashed string `H(EphID || i)` where `EphID = LEFTMOST128( H( seed_i ) )` as above. The outer hash-function is needed for security. The backend publishes the filter. All smartphones download it.

Each smartphone uses the filter `F` to check if in the past (i.e., before the corresponding filter `F` was published), it has observed any of the `EphIDs` reported by COVID-19 positive users. The phone checks if any of its stored hashes are included in the filter `F`.

For each matching beacon (e.g., a beacon with a recorded hashed identifier that was transmitted by a user who reported a positive diagnosis), the beacon's receive time and exposure measurement are provided to the exposure risk computation ([Section 4](#)).

Cuckoo filters have a low, but non-zero, probability of false positives, that is reporting that they contain an element that was not in the input set. In order to avoid unnecessarily alarming users, we select the parameters of the Cuckoo filter such that false positives are highly unlikely to occur even with heavy usage of the system over a number of years. In the scalability calculation below, we configure the filter to produce one false positive in a million users over a period of 5 years.

The use of a Cuckoo filter hides the set of ephemeral identifiers of COVID-19 positive users from the general public. The system uses a Cuckoo filter in conjunction with inputs that are obtained from cryptographic hashes of random values (the seeds, concatenated with timestamps). The inputs to the filter are sparse in a large set, i.e., the set of all possible inputs (128-bit strings). These two factors makes enumeration attacks against the filter an unattractive attack vector for adversaries, while still making it possible for users who have observed particular ephemeral identifiers to check for their inclusion in the filter. Attacks that attempt to reverse the filter and directly recover inputs from values held in the filter do not result in exposure of ephemeral IDs because of the extra layer of hashing performed on ephemeral IDs before entering them into the filter.

Scalability

This design requires more bandwidth and storage than in the low-cost design. The backend needs to store Cuckoo filters containing the hashed identifiers of COVID-19 positive users during the contagious period. Smartphones regularly download new cuckoo filters containing the latest hashed identifiers of COVID-positive patients. This data is static and can therefore be effectively served through a content delivery network. The computational cost on the phone is likely smaller than in the low-cost design, as phones only need to do one lookup per stored hashed identifier per cuckoo filter sent by the backend.

## 2.3. Hybrid decentralized proximity tracing

In this section, we present a hybrid design that combines ideas from the low-cost design and the unlinkable design. In this design, phones generate random seeds for each time window (for example, of length 2 hours) and use these seeds similar to the low-cost design to generate ephemeral identifiers for all epochs within that time window. Users upload seeds only if they are relevant to exposure estimation by other users.

Depending on the length of the time window, this design offers much better protection against linking ephemeral identifiers of COVID-19 positive users than the low-cost design and enables a user to redact time windows. The protection against tracking is weaker than the unlinkable design, but this scheme has a smaller bandwidth requirement.

This design is very similar to the Google/Apple design.[11] The Google/Apple design uses one seed to generate the ephemeral identifiers of that day, and thus corresponds to the special case where windows are 1 day long. In that configuration, the advantages with respect to the low-cost design are smaller. We recommend a time window of 2 or 4 hours depending on the bandwidth availability in the region.

Setup

No setup is needed.

---

## Generating ephemeral IDs

As in the previous designs, smartphones broadcast `EphID` during an epoch of fixed duration `L`. We group consecutive epochs into a time window `w`. The length of a time window can range from 10 minutes to a full day and needs to be an integer multiple of `L`.

At the start of each time window `w`, smartphones pick a new random 16-byte seed $\text{seed}_w$. Given the seed $\text{seed}_w$ for window `w`, each device computes

$$\text{EphID}_{w,1} \; || \; ... \; || \; \text{EphID}_{w,n} = \text{PRG( PRF(seed}_w\text{, "DP3T-HYBRID") ),}$$

where `PRF` is a pseudo-random function (e.g., HMAC-SHA256), "DP3T-HYBRID" is a fixed and public string, and PRG is a pseudorandom generator (e.g. AES in counter mode) producing `n * 16` bytes, which we split into 16-byte chunks to obtain the `n` ephemeral Bluetooth identifiers `EphID` for the window `w`.

Smartphones pick a random order in which to broadcast the `n` ephemeral Bluetooth identifiers within the window `w`. Each $\text{EphID}_i$ is broadcast for `L` minutes.

## Local storage of observed EphIDs

Smartphones locally record the observed beacons, similar to the low-cost design. For each received beacon the phone stores:

- The received ephemeral Bluetooth identifier `EphID`,

- The exposure measurement,

- The time window `w` in which the `EphID` was received.

For efficiency of storage, we propose to group these entries by EphID, resulting in 36 bytes per `EphID`. Given a very conservative estimate of observing 140k different `EphID`s over the course of 14 days (i.e., if epochs are 15 minutes, this would be 100 people observed per epoch), this would require 4.8 MB of local storage.

## Decentralized proximity tracing

In case of a positive diagnosis, users can instruct their device to upload the relevant seeds $\text{seed}_w$ generated by the smartphone during the contagious period. If the phone did not observe any `EphID` sufficiently close to be considered as an exposure during a time window `w`, it does not upload the corresponding seed $\text{seed}_w$ for efficiency. As in the unlinkable design, the user additionally has the option to **redact** identifiers, by choosing the set `W` of windows for which they want to reveal their identifiers. For example, the user may want to exclude windows for Monday morning and Friday night. The phone then uploads the set `{(w, seed`$_w$`)}` for all windows `w` in `W`.

The backend collects pairs `(w, seed`$_w$`)` of COVID-19 positive users. Phones periodically download these pairs. Each smartphone uses these pairs to reconstruct the list of

`EphIDs` of COVID-19 positive users for each window `w'` and checks if it has observed any of these `EphIDs` during window `w'` *in the past* (i.e., before the corresponding seed `seed`$_w$ was published). Restricting the matching to a specific time window limits replay attacks in which malicious users redistribute captured `EphIDs` and ensures more efficient lookups.

For each matching recorded beacon (e.g., a beacon with an `EphID` that was transmitted by a COVID-19 positive user), the beacon's receive time and exposure measurement are given as inputs to the exposure risk computation, see [Section 4](#).

## Scalability

This design requires more bandwidth and storage than the low-cost design, but less than the unlinkable design. The backend needs to store the `(w, seed`$_w$`)` pairs corresponding to COVID-19 positive users. Smartphones regularly download all new pairs. This data is static and can therefore be effectively served through a content delivery network.

The download cost depends on the length of the window and how many windows can be automatically omitted by the smartphone when uploading seeds. See Figure SH for a comparison. See the next section for how we computed these numbers.

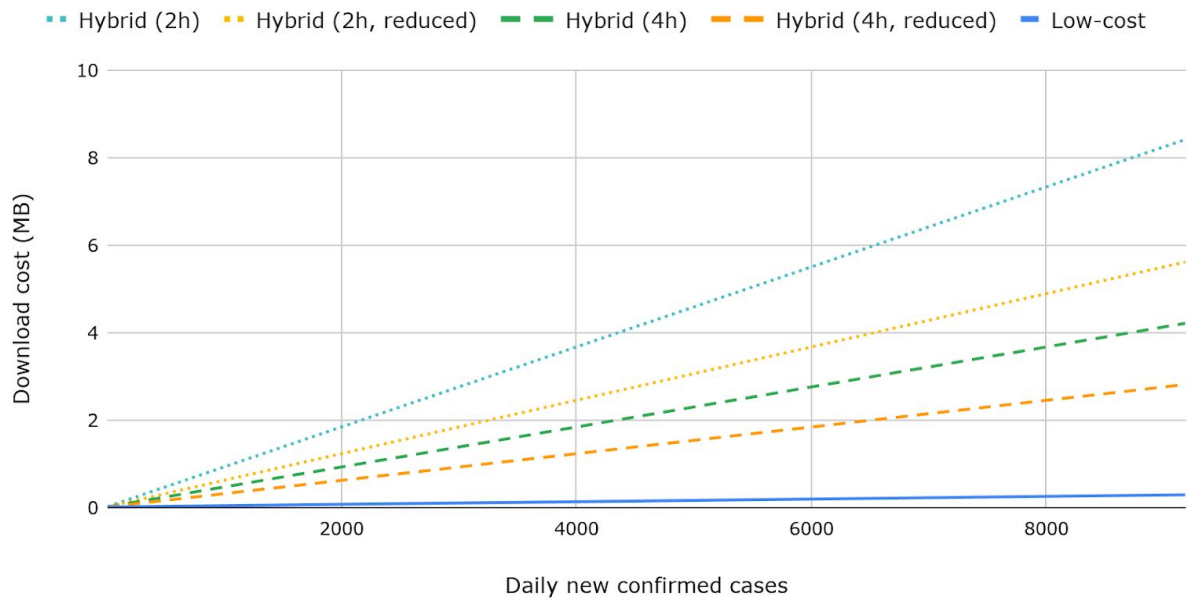

**Figure SH: Scalability of the hybrid design.** Comparison of the daily download cost per user (MB) depending on the number of new confirmed cases per day for different upload configurations of the hybrid design. We compare the download cost for different lengths of the time window `w` under two different assumptions. In the "normal" case, COVID-positive users upload seeds for all windows. In the "reduced" case their smartphone automatically omits the seeds for windows with a total length of 8 hours (e.g., because they were alone during that time and the phone did not detect any contacts).

## 2.4 Scalability

All three designs benefit from the use of a content delivery network (CDN). Smartphones of COVID-positive patients upload a small amount of data to the backend server. The backend server regularly redistributes this data to all other smartphones using a CDN. The daily download size scales linearly with the number of COVID-positive patients in all three designs. We assume a contagious window of 5 days and 15-minute epochs.

- In the low-cost design, the server needs to distribute one ($\texttt{SK}_t$, $\texttt{t}$) pair per diagnosed patient. This requires 36 bytes per patient.

- In the unlinkable design, the server needs to distribute 5 * 96 hashed strings per diagnosed patient. When using a well-tuned Cuckoo filter, this requires 2880 bytes per patient.

- In the hybrid design, we assume windows of 2 to 4 hours. The server must therefore distribute 5 * 6 or 5 * 12 seeds per diagnosed patient. This requires 480 to 960 bytes per user. Users that can redact 8 hours of windows only need to send 5 * 4 and 5 * 8 seeds, requiring 320 and 640 bytes respectively.

See Figure DC for the resulting daily download cost for smartphones. Table DC shows specific values based on peak rates in several countries.

We expect that proximity tracing systems, even when deployed in large EU countries, will operate in the range of up to 2000 new cases a day. At the time of writing, all large EU countries see less than 1500 new cases a day. We expect that if the rate of new cases increases, countries will take policy measures to restrict the infection rates. Thus, we expect the download cost to never exceed the single-digit requirement.

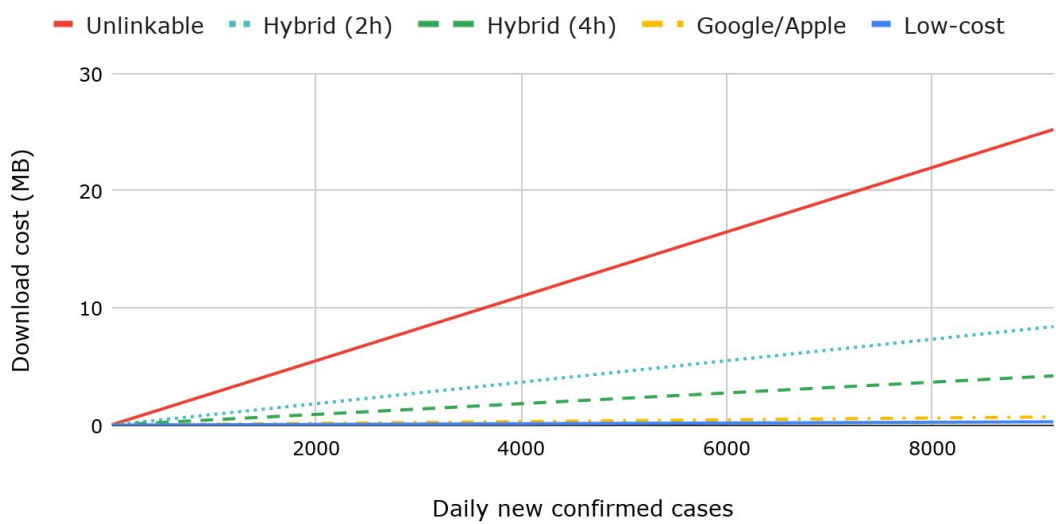

**Figure DC: Comparison of the daily download cost per user (MB) by number of new confirmed cases per day for different decentralised proximity tracing designs.**

**Table DC: Comparison of the daily per user download (MB) for different designs and countries.** For each country, we give the daily per user download at the overall peak of contagion and the maximum number of daily new cases between May 13 and May 20, 2020.

| | Low-cost (MB) | Unlinkable (MB) | Hybrid (4h) (MB) |
|---|---|---|---|
| **Switzerland (8M)** | | | |
| 1390 cases | 0.04 | 3.82 | 0.64 |
| 58 cases | 0.00 | 0.16 | 0.03 |
| **Germany (83M)** | | | |
| 6294 cases | 0.19 | 17.29 | 2.88 |
| 933 cases | 0.03 | 2.56 | 0.43 |
| **France (67M)** | | | |
| 7578 cases | 0.23 | 20.81 | 3.47 |
| 708 cases | 0.02 | 1.94 | 0.32 |
| **Spain (42M)** | | | |
| 9181 cases | 0.28 | 25.22 | 4.20 |
| 849 cases | 0.03 | 2.33 | 0.39 |
| **Italy (60M)** | | | |
| 6557 cases | 0.20 | 18.01 | 3.00 |
| 1402 cases | 0.04 | 3.85 | 0.64 |

## 3. Interoperability in decentralised proximity tracing systems

Effective proximity tracing systems must be interoperable across borders. Phones of users visiting foreign countries, whether it is for work, or for leisure, must be able to capture beacons from users in countries that they visit and include beacons of COVID-19 diagnosed patients in those countries in their exposure computation. Likewise, residents of a country must be able to receive notifications if a visitor to their country is diagnosed with COVID-19.

All three proposed designs support interoperability between different operators of different regions. Interoperability is possible as long as these operators use one of the decentralized designs proposed in this document. To interoperate with a specific protocol, the smartphone application must be able to process tracing data published for that protocol. That is, the application must run as many protocols as it needs to interoperate with.

To enable cross-border interoperability, backend servers of different regions (e.g., countries, or states) must exchange data. We propose the following mechanisms, explained in detail in other documents.[12]

First, when visiting a region, users enter the region into their phone. If a user visits a region frequently, e.g., workers commuting across borders, both regions can be permanently added to their phones. Phones use the list of visited regions to retrieve any proximity tracing data published by that region's backend for up to 14 days after the end of the users' visit. The way in which proximity data are published differs by protocol. For the low-cost design, this is a list of $(\mathtt{SK_t}, \mathtt{t})$ pairs, for the unlinkable design this is the cuckoo filter, and for the hybrid design this is the list of $(\mathtt{w}, \mathtt{seed_w})$ pairs. The phones then follow the protocol-specific procedures to match observed $\mathtt{EphIDs}$, which they then feed into the exposure risk computation described in the next section.

Second, to ensure all contacts of diagnosed users are notified, when a user is diagnosed with COVID-19, the phone will ask the user for recently visited regions. When uploading the seeds to the server, the phones also supply the list of visited regions. The backend authenticates the upload and then redistributes the uploaded tracing data to all visited regions. As a result, users in those regions will download this data and can determine if they observed any of the visitors' $\mathtt{EphIDs}$.

## 4. Exposure estimation

The goal of the exposure estimation is to estimate the duration of the smartphone owner's exposure to COVID-19 positive users in the past. This measurement serves as a proxy for the level of exposure to the SARS-CoV-2 virus. Local health authorities determine the exposure threshold for when a user should receive a notification. A prolonged exposure to the virus does not imply that the virus has been transmitted. However, the notification serves as a trigger for precautionary interventions recommended by local health authorities, such as testing or quarantine.

To compute exposure, the smartphone proceeds as follows. First, if necessary, it downloads the latest parameters provided by the health authority. Next, it takes all matches reported by the proximity tracing system for the past 14 days and estimates the exposure. In Switzerland, for instance, for each day, the phone combines the exposure measurements (e.g. signal attenuations) of all matches corresponding to that day, to compute a per-day exposure score.[13]

---

[12] For an overview see "Interoperability of decentralized proximity tracing systems across regions" retrieved from https://drive.google.com/file/d/1mGfE7rMKNmc51TG4ceE9PHEggN8rHOXk on 20 May 2020. A more detailed specification is provided here: "Decentralized Proximity Tracing: Interoperability Specification". The DP-3T Team (18 May 2020). Retrieved on 20 May 2020 from: https://github.com/DP-3T/documents/blob/master/DP3T%20-%20Interoperability%20Decentraliz ed%20Proximity%20Tracing%20Specification%20(Preview).pdf

[13] Details on the exposure estimation from BLE proximity measurements will be provided soon as separate documentation

If the exposure score is above the threshold determined by the health authority, the smartphone displays a notification that the user has been exposed to the virus through prolonged physical proximity to COVID-19 positive individuals. The notification advises the user on what to do and where to find more information. The details of the messages displayed, including the rate of repeated notifications and their content, need to be designed in close collaboration with health authorities and mental health experts.

## 5. Security and privacy considerations

In this section, we analyse the privacy and security properties of the three decentralised proximity tracing protocols introduced in this document. We have published a separate, far more extensive, risk evaluation of digital proximity tracing systems[14] that includes the class of decentralised systems our three designs belong to.

### 5.1 Threat model

In this section, we describe the adversaries that we take into account when carrying out the security and privacy analysis. For each of these adversaries, we describe their capabilities and the kind of risk they pose for the system. In the next section, we analyze the security and privacy of the system with respect to these adversaries.

**Regular user.** A typical user of the system that is assumed to be able to install and use the application by navigating its user interface (UI). They exclusively look at information available via the app's UI to infer private information about other users.

**Tech-savvy user** (Blackhat/Whitehat hacker, NGOs, Academic researchers, etc.)**.** This user has access to the system via the mobile App. In addition, she can set up (BT, WiFi, and Mobile) antennas to eavesdrop locally. Finally, she can decompile/modify the app, and she has access to the backend source code.

- (Whitehat hacker) Will investigate the App code, the information in the phone, and will look at what information is exchanged with the server (using an antenna or software installed on the phone, e.g., Lumen) or broadcast via Bluetooth (passive).

- (Malicious) Can DOS the system (targeted or system-wide), deviate from protocols, and actively broadcast Bluetooth identifiers.

**Eavesdropper** (Internet Service Provider, Local System administrators, Bluetooth sniffer)**.** They can observe network communication (i.e., source and destination of packages, payload, time) and/or BLE broadcast messages.

---

[14] "Privacy and Security Risk Evaluation of Digital Proximity Tracing Systems", The DP-3T Project, https://github.com/DP-3T/documents/blob/master/Security%20analysis/Privacy%20and%20Security%20Attacks%20on%20Digital%20Proximity%20Tracing%20Systems.pdf

- (Network adversary) Can use observed network traffic to attempt to determine the state of a user (e.g., whether they are at-risk, COVID-19 positive, etc.).

- (Local Bluetooth BLE Sniffer) Can observe local Bluetooth broadcasts (possibly with a powerful antenna to cover a wider area) and try to trace people.

It should be noted however that in many instances, for individuals or companies to use data in this way, or to collect data about passers-by to try and estimate their infection status based on the announced identifiers, will fall foul of a range of existing national and European laws around data protection, ePrivacy and computer misuse.

**Health authority.** Receives information about COVID-19 positive users as part of their normal operations to diagnose patients. The health authority learns information about at-risk people only when these at-risk people themselves reach out to the health authority (e.g., after receiving a notification from their app).

**Backend and App developers.** Can access all data stored at the servers. Moreover, the backend can query data from the mobile app in the same way that it would do during normal operations (in our designs, it can only change the data downloaded by the smartphones). They could also change the code of their backend software and the code of the mobile apps (including parameters related to proximity tracing). We assume they will not modify the mobile app because such action would be detectable. They can combine and correlate information, request information from apps, combine with other public information to learn (co-)location information of individuals.

**State-level adversary** (Law enforcement, intelligence agencies, etc)**.** Has the combined capabilities of the tech-savvy user and the eavesdropper. In addition, a state-level adversary can obtain subpoenas that give them the capabilities of the health authority, or the backend. Their goal is to obtain information about the population or to target particular individuals. They may be interested in past information, already stored in the system, or future information that will enable them to trace target individuals based on observed `EphIDs`.

**Unlimited-budget adversary.** An adversary with an unlimited budget, e.g., large organizations and (foreign) nation states, has the capabilities of tech-savvy users but can deploy these at a much larger scale. Additionally, such an adversary might be able to gain control over the project's infrastructure such as the backend. The goals of this adversary might be to learn information about the population or individuals (cf. a state-level adversary) or to disrupt the proximity tracing system, resulting in a form of denial of service. One form of disruption is to try cause a sizable part of the population to receive fake at-risk notifications by deploying antennas in public locations (e.g., airports, train stations, shopping malls, or parliament buildings) and relaying messages far and wide. This relay attack increases the chances of "at risk" contacts for the targeted population because their phones will perceive proximity where there is none.

## 5.2 Privacy

### *Privacy concerns*

**Social graph.** The social graph describes social relationships between users. Each node in the graph represents an individual user and an edge connecting two nodes indicates that there is a social relationship between the two users. A proximity tracing system does not need to provide information on the social graph to any party to fulfill its primary purpose.

**Interaction graph.** The interaction graph reflects close-range physical interactions between users. A labelled edge indicates an interaction between two adjacent users at a specific time. Knowledge of this graph is not necessary for proximity tracing nor for analyzing the spread of SARS-CoV-2. Therefore, *no party* needs to learn the interaction graph.

**Location traceability.** To perform proximity tracing, location traces (e.g. GPS coordinates) are not required. Therefore, no party in the system needs to have access to them or be able to easily trace individuals based on the BLE signals that the app broadcasts.

**At-risk individuals.** At-risk individuals are people who have recently been in contact with somebody who has tested positive for COVID-19. At-risk individuals need to know that they have been exposed to the virus so that they can take appropriate measures. No other party in the system needs to learn this information, other than when the notified user contacts and identifies herself to the health authority.

**COVID-19 positive status.** Only the user and the health authority need to know that the user has tested positive for COVID-19. No other party in the system needs to learn this information. In particular, app users do not need to know which of the individuals with whom they have been in contact have tested positive.

**(Highly) Exposed locations.** The system does not need to reveal any information about the locations that COVID-19 positive individuals have visited or the number of positive cases that have visited a specific location (e.g., to build a heat map of exposures). Proximity tracing can be performed without any party learning this information.

### *Privacy analysis of low-cost design*

**Social graph.** The low-cost design does not reveal any information to any entity. Any two users involved in a contact may learn this contact's existence from the system, but this was already known to them.

**Interaction graph.** The system does not reveal any information about the interaction between two users to any entity. The `EphIDs` revealed by COVID-19 positive users do not allow any inference about the people they have been in contact with to anyone except those contacts. The system thus prevents outside parties from learning the interaction graph.

**Location traceability.** In our low-cost design, the `EphIDs` of all users are unlinkable, and only the smartphone that generated them knows the corresponding seeds $SK_t$. When the phone's owner is diagnosed with SARS-Cov-2 and gives permission, the phone publishes to the backend the seed $SK_t$ corresponding to the first contagious day. After disclosing this information, the phone will generate a new seed at random. Given the seed $SK_t$ of the first contagious day, the `EphIDs` of a COVID-19 positive user are linkable from the start of the contagious window until the time of upload (at which point the phone picks a new seed).

As a result, tech-savvy users, eavesdroppers, and state-level adversaries can *locally* track infected patients during the (past) window in which the identifiers broadcasted via Bluetooth are linkable. To do so, the attacker uses strategically placed Bluetooth receivers and recording devices to receive `EphIDs`. The app's Bluetooth broadcasts of non-diagnosed users and COVID-19 positive users outside the contagious window remain unlinkable.

**At-risk individuals.** The seeds revealed to the server by COVID-19 positive users *are independent of* their contacts, i.e., the people they interacted with. They therefore do not give any information about people at risk to any party other than the at-risk individuals themselves.

**COVID-19 positive status.** Any proximity tracing system that informs a user that she has been in contact with a confirmed positive case inherently reveals a piece of information to the person at risk: one of the people they interacted with has tested positive for COVID-19.

A curious or malicious adversary might attempt to exploit this and other information in the system to identify the COVID-positive individuals with whom they have been in close proximity.

A curious user who only uses the standard interface of the app, cannot learn which of their contacts has tested positive because the app in normal operation does not reveal any information other than that the user was exposed at some point in the past. Such a curious user can only learn which of their contacts has tested positive if they learn this fact on an out-of-band channel (e.g., the COVID-positive person informs them, they observe the person going to the hospital, a common friend reveals the COVID-positve status, etc.).

A proactive tech-savvy adversary can abuse any proximity tracing system to identify individuals who have reported a positive diagnosis to the system and that she has been in close proximity with. This risk is a consequence of the basic proximity tracing functionality. The attack can be executed regardless of implementation and proximity tracing protocol (BLE or otherwise). It only relies on the single bit of information that any proximity tracing system must reveal — whether you have been exposed to a confirmed COVID-19 positive case.

To reveal an individual's COVID-19 positive status, the adversary must (1) keep a detailed log of who they saw and when, (2) register many accounts in the proximity tracing system, and (3) use each account for proximity tracing during a short time window. When one of these accounts is notified, the attacker can link the account identifier back to the time-window in which the contact occurred to learn when she was in close proximity to an individual who reported a positive diagnosis. The attacker can correlate this information with their detailed interaction log to narrow down who in their list of contacts is COVID-19 positive. In some cases, the adversary might even be able to single out an individual. This attack **is inherent to any proximity-based notification system,** as the adversary only uses the fact that they are notified together with additional information gathered by their phone or through other means.[15]

In decentralized proximity tracing systems, such as the three designs we propose in this white-paper, tech-savvy adversaries can learn when they were in close proximity to a COVID-19 positive individual without having to create multiple accounts. To determine when they interacted with a COVID-19 positive individual, they **proactively modify the app** [16] to store detailed logs of which EphID they received and when, and cross reference this list with the EphIDs reported by COVID-19 positive cases downloaded from the backend server. They then correlate exposure times with their log of who they saw to reveal which individuals they have been in contact with reported a positive diagnosis.

The low-cost design allows an adversary to **link** the EphIDs reported by COVID-19 positive cases, i.e. to learn which Bluetooth identifiers belong to the same device. COVID-19 positive individuals upload a single seed $SK_t$ that enables others to reconstruct, and thus link, a person's EphIDs for the entire contagious period. Due to the linkability of reported EphIDs, an attacker can **combine observations** at different times to identify who reported a positive diagnosis to the system. For example, the attacker might learn that the infected person she saw at 10:11AM is *the same* as the one she saw at 14:14PM. While she may have encountered many different people at each time, the intersection might be much smaller. This further increases the likelihood that the attacker can successfully single out a COVID-19 positive individual.

Tech-savvy users can also conduct a retroactive attack in which they attempt reidentification based on linkage and stored data, without the need to collect additional information in advance. The retroactive attacker only uses information stored by the app and auxiliary knowledge about the whereabouts of individuals during the contagious period. The data stored in the app provides coarse timing information when a specific

---

[15] For further details on this attack see "Privacy and Security Risk Evaluation of Digital Proximity Tracing Systems", The DP-3T Project,
https://github.com/DP-3T/documents/blob/master/Security%20analysis/Privacy%20and%20Security%20Attacks%20on%20Digital%20Proximity%20Tracing%20Systems.pdf
[16] We note that in some schemes such modifications would preclude the App from accessing measurement data entirely when using the Google and Apple API.

`EphID` has been observed, e.g., per day in the low-cost design. A tech-savvy adversary could leverage this information to single out a COVID-19 positive individual based on matching observed `EphIDs` to background knowledge of whom the adversary was with during this time window. A combination of multiple time windows might be enough to uniquely identify to whom the reported `EphIDs` belong. However, since smartphones broadcast the daily set of `EphIDs` in random order, the attacker cannot use the published seeds $SK_t$ to narrow down this coarse time-window. This decreases the likelihood that she will be able to successfully identify the COVID-19 positive individual in her contact list.

To **re-identify** an individual who has reported a positive diagnosis for COVID-19 to the system, an adversary needs to be able to **associate an identity** to the auxiliary information they have collected. For instance, a tech-savvy adversary who collects a detailed log of who they saw when needs to associate identities to each log entry. Without knowing the identities, the adversary cannot learn who tested positive. We can divide individuals the adversary interacts with into three groups depending on whether she will be able to reveal their identities or not:

- *Close individuals*: Family, friends, or colleagues with whom the adversary spends long periods of time. If these people received a positive diagnosis, they *will inform the adversary personally about their diagnosis* if they have spent time together. It is common practice that the authorities ask COVID-19 patients to notify any contact person at risk they remember.

- *Routine-sharing individuals*: People who share an activity with the adversary, such as riding a bus every day, supermarket tellers, etc. COVID-19 positive individuals in this group will likely not remember having been in contact with them and therefore will not (and cannot) notify the adversary.

- *Anonymous individuals:* People that the adversary sees sporadically and whose identities are unknown to the adversary.

As close individuals will reveal themselves, there is no extra information that an adversary can gain about the COVID-19 positive status of this group by exploiting the app. Anonymous COVID-19 positive users cannot be easily identified. Their privacy is only at risk if the adversary deploys additional (costly) means to associate identities with collected background knowledge. For instance, the adversary could attempt to combine data from surveillance cameras with facial recognition techniques to learn who is whom. The main group that is thus at risk through identification attacks is routine-sharing individuals.

We stress that in any case, having been close to an COVID-19 positive person is not proof of causality regarding transmission of the virus. Moreover, it is worth noting that reidentifying individuals and inferring their health status as a private entity without their permission would likely violate data protection law and, potentially, computer misuse law, which would further increase the cost and risk of undertaking this attack.

The pattern associated with the upload of identifiers to the server would reveal the COVID-19 positive status of users to network eavesdroppers (ISP or curious WiFi provider) and tech-savvy adversaries. If these adversaries can bind the observed IP address to a more stable identifier such as an ISP subscription number, then they can de-anonymize the confirmed positive cases. This can be mitigated by using dummy uploads. These dummy uploads provide plausible deniability to actual users' uploads, i.e., given an upload an observer cannot distinguish if it corresponds to an actual positive or a dummy. To avoid revealing which uploads were dummies to an adversary that polls the backend to learn if the list of ephemeral identifiers was updated, the backend should batch updates and only publish them in designated download slots.[17]

The backend server learns the IP address of COVID-19 positive users when they upload (a representation of) their `EphIDs`. If this adversary can bind the observed IP address to a more stable identifier, they can de-anonymize the confirmed positive patients. To reduce the risk, we recommend that the backend not log IPs.

*Mitigations*. In the current setting, retroactive attackers can link beacons received at different, coarse times to aid in identifying COVID-19 positive users. The amount of information available to such an attacker can be reduced by running the proximity tracing protocol either inside a privileged OS-level module[18] or inside a local trusted execution environment (TEE). These approaches isolate the proximity tracing protocol and the data they collect from users and malicious apps. The protocols running in the isolated environment would only output for each matching beacon: the corresponding exposure measurement (e.g., the attenuation) and a coarse time. The app then computes the exposure score (see [Section 4](#)). As a result, retroactive attackers can only learn the number of beacons of infected patients received each day but can no longer link beacons by the same COVID-19 positive patient.

By itself, this approach does not protect against tech-savvy users that proactively modify their device to collect beacons and then compute matching COVID-19 positive beacons using the public list of COVID-19 positive `EphIDs`. However, when using TEEs to isolate the proximity protocol, the system can be extended to hide this public list from tech-savvy users, ensuring that they **cannot recognize** COVID-19 positive beacons. To protect against tech-savvy users when using TEEs, the backend encrypts the list of seeds so that this list can only be decrypted *inside* the TEE. Each TEE downloads and decrypts the list of infected `EphIDs` and finds the matching beacons by cross referencing the list of infected `EphIDs` with the collected beacons. The TEE then returns to the app, for each day, a vector of the exposure measurements that enable the app to determine the user's exposure. As long as the TEE remains secure, tech-savvy users do not learn the `EphIDs` of COVID-19 positive patients.

Modern phones are equipped with TEEs that are used to harden smartphone kernels against attacks and to store cryptographic seeds. TEEs require buy-in from mobile platform providers (Apple, Google) and, for Android, the device manufacturers (Samsung, Huawei, etc.). The TEEs are well protected and difficult to attack even for tech-savvy users. While it is not impossible to leak this information, it is unlikely. We think such a mitigation is worthwhile in a later version of the proximity tracing system to further increase privacy guarantees. Other mitigation techniques could include the use of Private Information Retrieval and Private Set Intersection techniques, although current implementations may bring severe performance penalties.

---

[17] Details on the generation of dummy traffic will be provided in future documentation.
[18] This is the approach taken by the Google/Apple API.

Such technical measures as well as non-technical measures (e.g., banning modified applications from the market) could be introduced in case that the identification of COVID-19 positive individuals would become a threat to the system operation and to the users. The introduction of such measures depends on the overall risk assessment.

Finally, we note that if a small, cautious or misinformed portion of the population is concerned with these attacks and decides not to participate, this will not greatly impair the effectiveness of the deployment. As long as a large fraction of the population runs the app, the number of at-risk identifications will be large enough to significantly reduce the rate of transmission.

**(Highly) Exposed locations.** A powerful tech-savvy adversary operating its own BLE equipment from a single location can collect `EphIDs` within 20-100m range, depending on the phone output power and environment. When combining this list with the `EphIDs` that can be computed from the `SKs` downloaded to the phone, an adversary could learn whether any COVID-19 positive user has visited the location in a small radius of 50m. Furthermore, the adversary could reveal how many distinct diagnosed persons have visited the location in the past.

### *Privacy analysis of unlinkable design*

The unlinkable design provides overall better privacy properties at the cost of increased bandwidth. The two designs provide the same level of protection for the social and interaction graph. We address the remaining differences point by point.

**Location traceability.** In the unlinkable design, the `EphIDs` remain unlinkable for all users against a local attacker. This unlinkability also holds for COVID-19 positive patients so long as the server is honest. However, if the server is malicious, then it can infer which ephemeral identifiers belong to a COVID-19 positive user through timing information or other metadata created when ephemeral identifiers are uploaded to the server. The use of anonymous communications could mitigate this threat.

**At-risk individuals.** As in the low-cost design, the seeds revealed to the server by users who have received a positive diagnosis *are independent of* their contacts. Hence, they do not give any information about people at risk.

The rest of the analysis is the same as for the low-cost design.

**COVID-19 positive status.** The unlinkable design reduces the linkability of the `EphIDs` reported by a COVID-19 positive user. Compared to the low-cost design, this reduces the likelihood that a proactive or retroactive tech-savvy adversary can identify which of their contacts has reported a positive diagnosis through linkage attacks. The adversary can no longer combine observations at multiple points of time to single out a COVID-19 positive individual.

Retroactive attackers can extract little information from the records stored on the phone. For each beacon, the phone only stores the hashed string, a coarse receive time (e.g., the day on which the beacon was received) and the exposure measurement. Retroactive attackers cannot recover a more detailed receive time, and thus only learn the total count of COVID-19 positive beacons for each day.

However, a proactive tech-savvy adversary can still modify their application to learn when she has been in close proximity to a confirmed positive case. As described in our detailed privacy risk evaluation,[19] this attack can be executed in any proximity tracing system by using multiple accounts. It cannot be avoided.

The unlinkable design enables COVID-19 positive individuals to redact periods of time that they consider sensitive and for which they prefer not to disclose their contacts. This can alleviate concerns in a close-knit or small community in which users may be concerned that community members learn of their positive diagnosis through the app, instead of being informed in person.

**(Highly) Exposed locations.** As in any practical BLE-based PT system, an adversary could identify locations that have been visited by COVID-positive users in the past. However, as EphIDs cannot be linked to a single device, it is more difficult for the adversary to learn how many distinct cases visited the location.

### *Privacy analysis of hybrid design*

The hybrid design provides privacy properties similar to the low-cost and the unlinkable design. There are two major differences between the hybrid and the low-cost design.

1) The hybrid design controls the linkability of the ephemeral identifiers reported by COVID-19 positive users and restricts linkability to short to medium-length time windows.

2) The hybrid design allows users who report a positive diagnosis to redact identifiers for specific time windows before sharing them with other devices.

---

[19] See "Privacy and Security Risk Evaluation of Digital Proximity Tracing Systems", The DP-3T Project,
https://github.com/DP-3T/documents/blob/master/Security%20analysis/Privacy%20and%20Security%20Attacks%20on%20Digital%20Proximity%20Tracing%20Systems.pdf

These two differences affect the privacy properties of the hybrid design in the following ways:

**COVID-19 positive status.** Proactive and retroactive linkage attacks by tech-savvy adversaries aim to reveal the COVID-19 positive status of individuals they have been in close proximity with. To learn this information, the adversary needs to extract from the system *at which times* they have been in contact with a COVID-19 positive user. They can then correlate this information to auxiliary knowledge about who they saw when to identify individuals who reported a positive diagnosis. If the adversary can link ephemeral identifiers, i.e., associate multiple of the EphIDs reported by COVID-19 positive users to the same individual, the adversary can combine observations from different time frames to single out individuals. The hybrid design restricts the linkability of reported EphIDs to a time window `w`. This reduces the likelihood that the adversary can successfully narrow down the group of contacts who might have tested positive.

In comparison to the unlinkable design, the medium-term linkability of EphIDs implies the hybrid design provides slightly less protection against proactive and retroactive identification attacks. We note though, that as in all decentralized designs discussed in this white paper, the proactive tech-savvy user must modify their application to record receive times.

The hybrid design allows COVID-19 positive users to decide not to share their broadcast identifiers for certain time windows. This further reduces the likelihood of successful identification attacks as the adversary cannot use auxiliary information for the redacted time frames to reveal the identity of diagnosed users.

Retroactive attackers can extract some information from the records stored on the phone. For each beacon, the phone stores the `EphID`, a time and the exposure measurement. Because `EphIDs` from the same COVID-19 positive patient are linkable during a time window, the retroactive attacker can estimate contact duration with a single COVID-19 positive patient for each time window.[20]

**(Highly) Exposed locations.** An adversary is less likely to be able to learn the number of positive cases who visited a specific location because of the restricted linkability of ephemeral identifiers. The adversary can link EphIDs for the duration of a time window but cannot link identifiers of the same individual across multiple time windows.

## 5.3 Security

### *Security concerns*

**Fake exposure events.** A fake exposure event could make a person believe that they are at risk, even though they have never been exposed to a diagnosed user. Attackers could try to generate fake exposure events to trigger false alerts, e.g. by relaying or broadcasting EphIDs at large scale. This would violate the authenticity requirement of the system.

---

[20] This retroactive attack does not work when using the Google/Apple API as it does not expose received EphID to any application or user.

**Suppressing at-risk contacts:** There is a risk that either a COVID-19 positive user or the backend server could prevent other individuals from learning they are at risk, e.g., by modifying the app's local storage. This violates the integrity of the system and would lead to an increased health risk for at-risk individuals who rely on the system for alerts.

**Prevent contact discovery:** A malicious actor could disrupt the system, e.g. by jamming Bluetooth signals, and prevent contact discovery.

## *Security analysis of low-cost design*

**Fake exposure events.** In all practical proximity tracing systems based on Bluetooth-based exposure measurements, an adversary with a powerful antenna can trigger false alerts of an exposure to a COVID-19 positive person that do not reflect real-world proximity to a positive-case person.

To cause false alarms, a malicious adversary simply places her proximity tracing device in a crowded area and hooks up a powerful transmitter to *artificially increase the range* of her Bluetooth contacts. As a result, other devices located beyond 2 meters can interact with the attacker's device and will perceive the attacker's device as "near-by". To complete the attack, the attacker must ensure that these interactions between her device and other devices are flagged as exposure events. To do so, the attacker either:

1. **Herself tests positive** and brings her device to the hospital when she gets tested (requiring the adversary to be infected).

2. **Bribes a diagnosed person** to bring the attacker's device to the hospital instead of their own (or simply obtains the upload authorization code from them).

3. **Hijacks/bribes the health authority** that authorises COVID-19 positive individuals to trigger proximity tracing.

4. **Compromises the backend server** that sends information or directly notifies users of the system.

In the low-cost design, an attacker can record an individual's ephemeral identifier and broadcast it to victims at a different location and/or time, as long as it is **relayed on the same day**. If that individual later receives a positive diagnosis, the victims will incorrectly believe they have been exposed.

In the low-cost design, the seeds of COVID-19 positive users shared for exposure calculation are bound to the day on which they were valid. This prevents relay attacks in which the adversary attempts to relay an individual's ephemeral identifiers with a delay of more than 24 hours.

An attacker could be motivated to claim another user's `EphID` as their own and report that it should be included in the exposure risk calculation. The low-cost design addresses this risk by requiring users to upload the seeds $SK_t$ from which their `EphIDs` are derived. As these `EphIDs` are derived from the seed using a cryptographic hash function and a pseudo-random function, it is computationally infeasible for an attack to learn another user's seed from observing their broadcasts.

**Suppressing at-risk contacts.** Hiding at-risk contacts is possible in any proximity tracing system. Infected users can choose to not participate at all; to temporarily not broadcast Bluetooth identifiers, or not to upload their data once diagnosed.

**Prevent contact discovery.** Any proximity tracing system based on Bluetooth low energy is susceptible to jamming attacks by active adversaries. Such jamming attacks will cause the normal recording of `EphIDs` to stop working, hence preventing contact discovery. This is an inherent problem of this approach.

## Security analysis of unlinkable design

The unlinkable design has the same security properties as the low-cost design with respect to **suppressing at-risk contacts** and **preventing contact discovery**.

**Fake exposure events.** As in all practical proximity tracing systems based on BLE handshakes between personal smartphones, a powerful adversary can cause false alarms through BLE range extension attacks.[21]

In the unlinkable design, ephemeral identifiers are cryptographically linked to the *epoch* in which they are broadcast. To create fake exposure events, the attacker must therefore receive and rebroadcast `EphIDs` *within the same epoch*. Such an **"online" relay attack** is unavoidable in proximity tracing systems based on passive Bluetooth advertisements.

As in the low-cost design, the `EphID` generation protocol of the unlinkable design prevents an attacker from claiming another user's `EphID` as their own. To do so, an attacker would have to be able to infer a user's seed seedt from their broadcast identifier which is computationally infeasible.

## Security analysis of hybrid design

The hybrid design has the same security properties as the low-cost design with respect to the risks of suppressing at-risk contact and preventing contact discovery.

---

[21] For further details on this general attack see "Privacy and Security Risk Evaluation of Digital Proximity Tracing Systems", The DP-3T Project, https://github.com/DP-3T/documents/blob/master/Security%20analysis/Privacy%20and%20Security%20Attacks%20on%20Digital%20Proximity%20Tracing%20Systems.pdf

**Fake exposure events.** As in all practical proximity tracing systems based on BLE handshakes between personal smartphones, a powerful adversary can cause false alarms through BLE range extension attacks.

In the hybrid design, EphIDs are linked to the valid time window of the seed they were derived from. To create fake exposure events, the adversary must therefore receive and broadcast EphIDs within the same time window. This prevents relay attacks in which the adversary attempts to relay an individual's ephemeral identifiers with a delay of more than the length of a time window.

As in the low-cost design, the `EphID` generation protocol of the hybrid design prevents an attacker from claiming another user's `EphID` as their own. To do so, an attacker would have to be able to infer a user's seed value $seed_w$ from their broadcast identifier which is computationally infeasible.

## 6. Protection from short-term and remote eavesdropping at the physical layer

In this section, we introduce an enhancement to the decentralised proximity tracing solutions proposed in this document. It would also apply to similar initiatives such as PACT and TCN,[22] and the joint Apple and Google Exposure Notification protocol.[23]

A shortcoming in most decentralized proximity tracing systems based on the exchange of BT advertisements between devices is that a malicious party who is willing to modify their app or deploy their own software is able to record a proximity event ***despite only being in contact for a short amount of time or at a long distance***. This violates the requirement that the system provide ***precise*** data, i.e. only report exposure events that represent actual physical proximity.

In particular, an attacker could attempt to gather a significant number of `EphIDs` by deploying specialist equipment, either in high-traffic locations or in a vehicle that can cover a wide area ("wardriving"). The attacker can also deploy high gain, directional antennas to cover wide areas, further increasing the range and selectivity of the attack. The attacker can later see which of the recorded `EphIDs` correspond to users who reported a positive diagnosis and use additional metadata such as location, timing, video surveillance, etc. to infer their identities.

We note that for these enhancements to be efficient, changes in low-level smartphone components (e.g., Bluetooth chips) are likely necessary. We expect that without these changes, these enhancements will have a non-negligible impact on battery life.

---

[22]For PACT, see Justin Chan et al. (2020) PACT: Privacy Sensitive Protocols and Mechanisms for Mobile Contact Tracing, retrieved from: https://covidsafe.cs.washington.edu/on 20 May 2020. For TCN, see TCN Coalition (2020) TCN Protocol, retrieved from https://github.com/TCNCoalition/TCN on 20 May 2020.
[23]See https://www.apple.com/covid19/contacttracing/

### *EphID* spreading with secret sharing

To address these problems, we introduce an enhancement to our system: `EphID` *Spreading With Secret Sharing*.

In a nutshell, this enhancement spreads each ephemeral identifier `EphID` across low-power beacons using a ***k-out-of-n secret sharing scheme***. Instead of transmitting each `EphID` within a single beacon, we encode it into n shares, such that each receiver needs to receive at least k shares to reconstruct the `EphID`. There are a number of secret sharing techniques that could be used for this purpose. We are currently running experiments to determine which is the most robust scheme for the scenarios in which these systems are to be deployed.

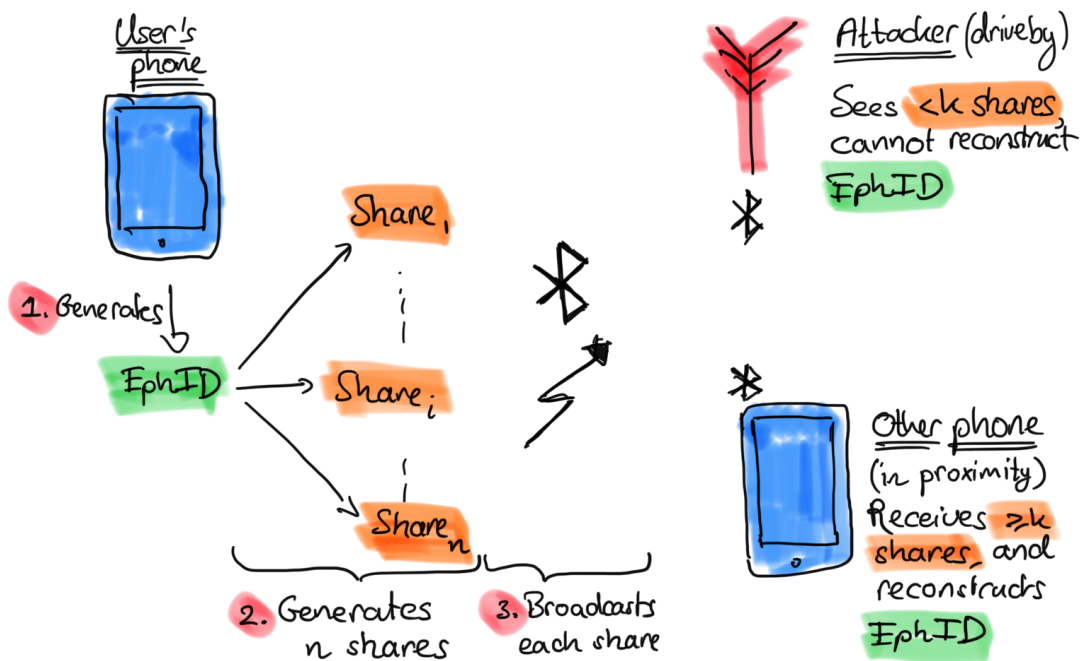

**Figure SW: `EphID` spreading with secret sharing.**

In our earlier designs, each device divides time into epochs and picks a random `EphID`$_i$ for each epoch `i`. The `EphID` is transmitted several times during the epoch and each broadcast contains the whole `EphID`. The broadcast frequency depends on manufacturer- and implementation-specific details.

With this enhancement, each broadcast contains shares of the `EphID`. If `EphID` is to be retransmitted within an epoch, new shares are generated. In the simplest case, however, the number n of shares can be set to equal the number of broadcasts that the device makes within an epoch. We stress that this enhancement, even with the spread of the `EphID`s, should not have a significant impact on battery life. It requires no additional transmission or reception over the basic designs, and the additional computation is minimal.

## Tuning the trade-off between privacy and utility

*Tuning privacy parameters.* Setting the value k requires careful consideration. A system must receive enough beacons during the contact interval (as determined by epidemiologists) to receive k shares and register a contact. A smaller k thus increases the robustness of the system in normal operation. However, the larger value of k, the longer an adversary is forced to shadow a victim in order to collect a sufficient number of beacons to reconstruct the `EphID`. Hence, the proposed number k of shares is a trade-off between privacy and the ability to record short contacts.

*Eavesdropping from a distance.* Spreading `EphID` across beacons not only protects against eavesdropping by adversaries who are only briefly collocated with their victim, but it also makes eavesdropping from a distance much harder. The requirement to successfully receive multiple broadcasts increases the asymmetry between a legitimate receiver in proximity and a malicious eavesdropper at a distance. An eavesdropper who is placed further away will typically experience a worse channel to the transmitter and a higher packet loss.

If we assume an attacker without access to specialised equipment and with reasonable assumptions on the broadcast transmission power, frequency, and probability of successful reception, we can select k, n such that a close by, legitimate user within 5 meters would have a very high probability of successfully receiving an `EphID` within a reasonable contact time threshold of five minutes (>99.9%), but an attacker attempting to eavesdrop from 16 meters away would have a small probability of success (<1%). An attacker using specialised hardware would be able to improve their odds either by increasing their probability of successful reception or by cryptographic analysis of the malformed broadcasts.

To achieve the appropriate balance between the desired range of reception of `EphIDs` (epidemiologically relevant) and the resilience to eavesdropping, we need to select the right combination of transmission power, transmission frequency, and required number k of reconstruction shares. We expect these parameters to be configurable and determined by further experiments, functional requirements, and risk assessment. The use of ultra-low or low power beacons will likely best protect the privacy of the users and facilitate proximity detection.

Our scheme can be integrated within a ranging technique or used in addition to existing (e.g., RSSI-based) ranging. In the latter case, ranging would use a different epoch identifier that is different but linked to the `EphID` that the device is broadcasting. If supported by BLE chipsets, this scheme could be further enhanced by in addition distributing the shares across three BLE advertisement channels.

# 7. Comparison with centralized approaches

We classify two key functionalities in proximity tracing systems that are decentralized in our schemes:

- Ephemeral identifier generation: Ephemeral identifiers broadcast via Bluetooth are generated on the phone.

- Exposure estimation: The estimation of the exposure is computed locally on the phone.

We now compare the security and privacy properties of the decentralized approaches presented above with schemes in which both operations are centralized, and with schemes in which seeds are generated on phones but COVID-exposure estimation is centralized.

## 7.1 Centralized identifier generation and exposure estimation

In approaches in which both identifier generation and COVID-exposure estimation are centralized, such as ROBERT,[24] PEPP-PT-NTK,[25] and OpenTrace/BlueTrace/TraceTogether,[26] a central server estimates a user's likelihood of COVID-exposure, instead of the user's smartphone in decentralized designs. Depending on the system, the server notifies the at-risk users (PEPP-PT-NTK, OpenTrace) or users query the server about their status (ROBERT).

In all these systems, the central server holds a long-term pseudo-identifier for every user and uses it to derive ephemeral pseudo-identities ($\texttt{EphIDs}$) that are pushed to the smartphones.

The smartphones broadcast the $\texttt{EphIDs}$ received from the central server and record the $\texttt{EphIDs}$ transmitted by near-by smartphones. Smartphones *locally store all* observed $\texttt{EphIDs}$ together with their corresponding proximity and duration. See Figure ZZ.

In case of a positive diagnosis, users can give permission for their smartphone to send the recorded list of observations to the server to enable proximity tracing.

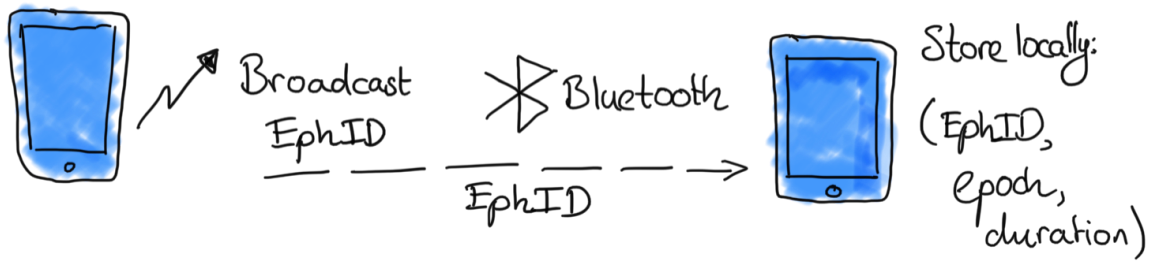

**Figure ZZ: Processing and storing of observed `EphIDs`.**

## Central proximity tracing

### PEPP-PT-NTK and OpenTrace

The PEPP-PT-NTK and OpenTrace backends execute the proximity tracing process after a diagnosed user has uploaded their list of observations [(`EphID`, epoch, duration)] for the contagious window. The backend recovers the long-term pseudo-identifiers of the at-risk users from the reported observed `EphIDs` and triggers a process to notify them if their exposure is high enough. See Figure ZY.

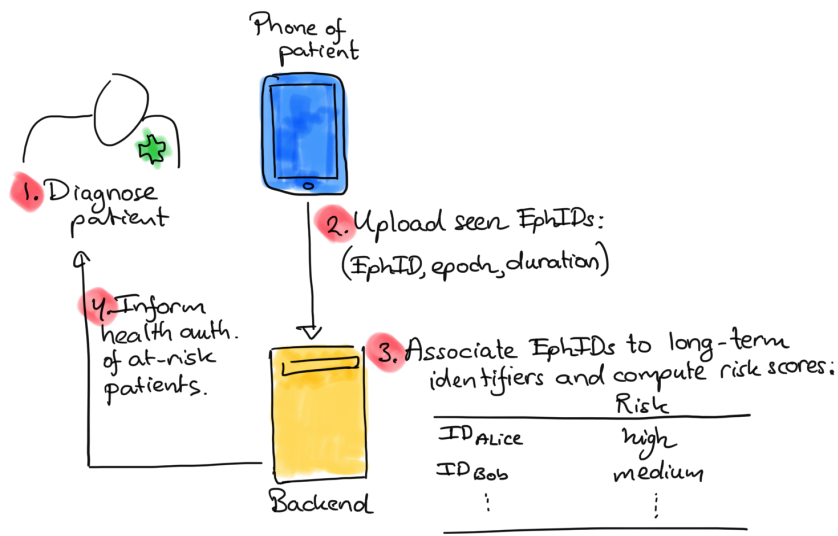

**Figure ZY: Proximity tracing for centralized PEPP-PT-NTK and OpenTrace designs.** In ROBERT, the users instead query the backend.

### ROBERT

In the ROBERT system, the backend tracks the exposure of each user of the system. As in PEPP-PT-NTK and OpenTrace, COVID-19 diagnosed patients upload their observed EphIDs to the backend server. The backend server associates these observed EphIDs to long-term pseudo-identifiers of at-risk users. The backend uses the associated data to update the exposure for each of the at-risk users.

Unlike PEPP-PT-NTK and OpenTrace, the backend does not notify patients. Instead, smartphones of ROBERT users regularly query the backend to request their exposure status. The backend answers whether the exposure passed the threshold or not.

## 7.2 Local identifier generation and centralized exposure estimation

Other approaches, such as DESIRE[27] instead generate identifiers on the phone, while still estimating COVID-exposure centrally. Instead of broadcasting self-contained, ephemeral identifiers, smartphones in DESIRE broadcast ephemeral public keys that, when combined with others' public keys, yield ephemeral identifiers `EphIDs`.

In case of a positive diagnosis, users can give permission for their smartphone to send a version of the observed `EphIDs` to the backend to enable proximity tracing.

### Central proximity tracing

Smartphones regularly query the backend to request their current exposure status. To enable the backend to compute this status, phones upload a version of the `EphIDs` computed from all the encounters they had in the relevant period. The server takes all observations reported of these encounters [(`EphID`, epoch, duration)] and estimates exposure. If the exposure is long enough, the user receives a positive exposure response. Otherwise, the user receives a negative.

## 7.3 Privacy comparison

**Social graph.** In a system in which both identifiers and COVID-exposure are computed centrally, the backend server can always associate ephemeral broadcast identifiers with permanent pseudo-identifiers for individual devices. If `EphIDs` are associated with a long-term identifier (e.g., in PEPP-PT-NTK), the backend server can reconstruct the social graph of users from the information shared by COVID-19 positive users. The server can join subgraphs from different positive cases to gain a comprehensive picture of the true underlying social graph. Given other partial social graphs with identities, the server can match its graph to the other graphs and reidentify nodes.

ROBERT and DESIRE propose to prevent the leakage of the social graph by using an anonymous communication network to upload observed identifiers in an unlinkable manner. In this way, the backend cannot associate uploads to a user nor determine which identifiers were observed by the same COVID-19 positive user. As a result, the backend cannot reconstruct an infected user's contacts.

---

[27] Castellucia et al. (9 May 2020) DESIRE: A third way for a European Exposure Notification System https://github.com/3rd-ways-for-EU-exposure-notification/project-DESIRE/blob/master/DESIRE-specification-EN-v1_0.pdf on 19 May 2020.

In separate documents, we show that (1) these mechanisms when applied to ROBERT are ineffective and still allow reconstruction of the social graph;[28] and (2) are difficult to realise in practice for DESIRE.[29]

**Interaction graph.** In a system in which both identifiers and COVID-exposure are computed centrally and uploaded observed identifiers are linked, the backend server can always associate uploaded ephemeral broadcast identifiers to permanent pseudo-identifiers for individual devices. In PEPP-PT-NTK, OpenTrace, and ROBERT, observed identifiers are timestamped. Thus the backend server can not only reconstruct a social graph, but it can reconstruct an interaction graph.

The subset of the full interaction graph learned by a server grows quickly as every newly confirmed positive user uploads their entire contact history, which can be linked to existing nodes in the graph. Even though the nodes in the graph are pseudonymous, this is a serious privacy concern because graph data is easy to reidentify.[28]

If deployed (including anonymous communication, ensuring enough mixing with uploads of other COVID-positive patients, and anonymous authentication), the mechanisms in DESIRE to protect the social graph preclude the DESIRE backend from learning the interaction graph.

**Location traceability.** The decentralized design limits the potential for location tracking to users who have received a positive diagnosis and for the course of the contagious period. In centralised systems in which keys are generated on the server, access to server-side keys (e.g., the backend itself or law enforcement) enables linking ephemeral `EphIDs` to the corresponding permanent app identifier. This enables tracing/identifying people based on `EphIDs` observed in the past, as well as tracing peoples' future movements.

When keys are generated on the phone and not used directly as `EphIDs`, as in DESIRE, location traceability is equivalent to the [decentralised unlinkable design](#).

**At-risk individuals.** In centralised systems in which the server controls key generation and notifies the user (PEPP-PT-NTK, OpenTrace), by design, the backend recovers the identity of an at-risk individual to notify these individuals.

In other centralised designs (ROBERT, DESIRE), users query the server to learn their exposure status. The identity of at-risk users is only protected when servers cannot deanonymize users through their permanent app identifiers and network identifiers.

As in decentralized designs, network eavesdroppers do not learn at-risk status.

---

**COVID-19 positive status.** The centralised and decentralised proximity tracing systems share the inherent privacy limitation that they can be exploited by a tech-savvy user to reveal which individuals in their contact list might be infected. However, the centralised designs hide when and how often the user was in contact with a COVID-positive patient. As a result, tech-savvy attackers cannot benefit from linking between `EphIDs` and timing information to amplify their attack. Instead, they need to rely on multiple accounts.

Depending on whether the centralized designs deploy dummy traffic correctly, network eavesdroppers might still learn the COVID-19 status of users of the system.

## 7.4 Security comparison

**Fake exposure events.** Triggering false alerts is easy in all centralised designs except DESIRE and can be done retroactively by any tech-savvy COVID-19 positive user. It does not require broadcasting. It suffices to add the target's `EphIDs` to the list of observed events prior to uploading them to the backend.

DESIRE requires an active exchange between users to trigger a fake exposure event, and therefore requires broadcasting.[30]

**Suppressing at-risk contacts.** Hiding at-risk contacts is possible in any proximity tracing system.

**Prevent contact discovery.** Any proximity tracing system based on Bluetooth BLE is susceptible to jamming attacks by active adversaries.

---

[30] For further details on this general attack see "Privacy and Security Risk Evaluation of Digital Proximity Tracing Systems", The DP-3T Project, https://github.com/DP-3T/documents/blob/master/Security%20analysis/Privacy%20and%20Security%20Attacks%20on%20Digital%20Proximity%20Tracing%20Systems.pdf

## 8. Conclusion

In this whitepaper, we designed a privacy-preserving proximity tracing system and analyzed three different protocols. All three protocols minimize exposure of private data, limiting the risk of a privacy leakage.

Our design relies on smartphones to *locally* compute the exposure of an individual user to the SARS-CoV-2 virus through proximity over a prolonged period of time to COVID-19 positive people. Data about specific exposure events, i.e., interactions of people, always remains on a user's phone.

The three implementations offer different trade-offs between bandwidth and privacy protection. One design results in an extremely lightweight system. The others offer extra privacy properties at the cost of a small increase in download data size. The three alternatives scale to a large number of users with minimal local computation and minimal centralization.

We also provided evaluation criteria to assess the level of privacy provided by any proximity tracing solution. We thoroughly evaluated our protocols with respect to performance, security, and privacy. Compared to central designs in which the backend computes risks and informs users, our design protects interaction graphs from the backend. Only a determined, tech-savvy adversary can learn any extra information besides that made visible by the app. The centralized system, in comparison, leaks to the backend unnecessary information about contacts and requires a large amount of trust in a central entity.

Our three implementations show that there are a wide range of alternatives to be explored among the trade-off between resistance to different active and passive attacks, battery consumption, and bandwidth overhead. We encourage researchers and technology companies working on proximity tracing to continue searching for the best realistic operation point.